\RequirePackage{ifpdf}
\documentclass{pCHEF}

\title{Prototype studies for a forward EM calorimeter in ALICE}

\ShortTitle{ALICE FoCal Prototype}

\author{\speaker{Thomas Peitzmann}\thanks{for the ALICE FoCal Collaboration}\\
        Utrecht University and Nikhef\\
        E-mail: \email{t.peitzmann@uu.nl}}

\abstract{A forward electromagnetic calorimeter (FoCal) based on SiW technology is being considered as a possible upgrade to the ALICE detector. This device should in particular feature an extremely high granularity allowing $\gamma/\pi^0$ discrimination out to very high momenta. The main option considered for the high granularity layers is the use of CMOS pixel sensors. We will discuss the motivation and design principles of the proposed calorimeter and will then focus on the experience gained with a full CMOS-pixel calorimeter prototype. Preliminary results of this device show very good capabilities for shower profile measurements, and reasonable results for linearity and resolution, which are limited by the currently incomplete calibration.}

\FullConference{Calorimetry for High Energy Frontiers - CHEF 2013,\\
		April 22-25, 2013\\
		Paris, France}

\usepackage{CHEF2013_definitions}
\newcommand{\JETPHOX} {{\textsc{jetphox}}\xspace}
\begin{document}

\section{Physics Motivation and Design Principles}

The ALICE experiment at the CERN LHC has an ambitious upgrade program \cite{Upgrade_LoI}, which has been endorsed by the LHCC. It includes upgrades of the rate capabilities and of its central tracking detectors to enable unique precision studies of Pb-Pb collisions. In addition, an enhancement of the forward muon measurements by additional tracking in front of the muon absorber will be proposed.
Furthermore, a calorimeter at forward rapidities (FoCal) is being discussed internally. This detector would be intended to measure photons, electrons/positrons and jets for rapidities $\eta > 3$. Such a detector would offer a wealth of physics possibilities, but its main focus is on measurements related to the structure of nucleons and nuclei at very low Bjorken-$x$ and possible effects of gluon saturation (often modelled as the so-called the colour glass condensate -- CGC) \cite{1994PhRvD..49.2233M,1994PhRvD..49.3352M,1994PhRvD..50.2225M}. The cleanest signals for such initial-state modifications are prompt photons, which are not modified by final-state strong interactions and are also produced with sufficient rates. 

\begin{figure}[h!] 
\begin{center}
\includegraphics[width=.6\textwidth]{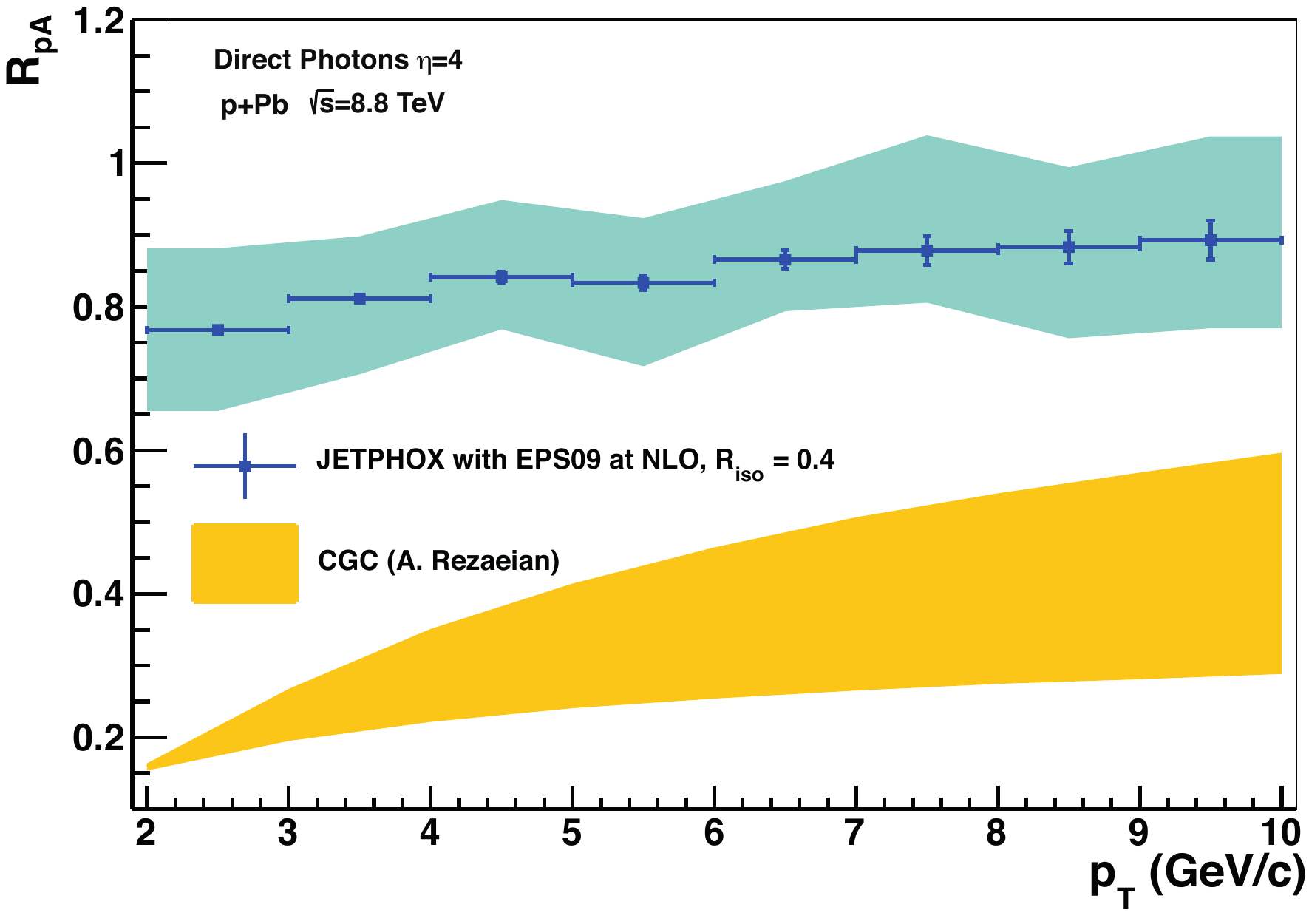}
\caption{Nuclear modification factor for direct photons as a function of \pt at $\eta = 4$ for p--Pb collisions at 8.8 TeV at the LHC. Predictions for the gluon saturation model from \cite{rezaeian2013} are compared to calculations for isolated photons in NLO pQCD using \JETPHOX \cite{1126-6708-2002-05-028} with the EPS09 PDF sets. The blue shaded area indicates the uncertainty associated with those PDFs, while the orange band is an estimate of the systematic uncertainty of the CGC calculation.} 
\label{fig-rpa} 
\end{center}
\end{figure}

Evidence for the CGC could be obtained e.g. by measuring the nuclear modification factor of isolated photons:
\begin{displaymath}
R_{pPb} \equiv =\frac{1/\pt \,\, dN/d\pt (pPb)}{\left\langle N_{\mathrm{coll}} \right\rangle \cdot1/\pt  \,\, dN/d\pt (pp)}.
\end{displaymath}
Predictions for $R_{pPb}$ from different theoretical calculations are shown in Fig.~\ref{fig-rpa}. An NLO pQCD calculation with the  \JETPHOX program \cite{1126-6708-2002-05-028} using EPS09 nuclear PDF sets shows a moderate suppression by at most $20 \%$ related to nuclear shadowing, while the photons are strongly suppressed ($\approx 80 \%$) at low \pt in the CGC calculation.

FoCal would consist of an electromagnetic calorimeter most likely positioned at a distance from the IP of $z \approx 7 \,\mathrm{m}$ covering $3.2 < \eta < 5.3$ backed by a standard hadronic calorimeter. A distance of $z = 3.6 \,\mathrm{m}$, which corresponds to a maximum reachable pseudorapidity of $\eta = 4.5$, has also been studied in simulations. Both positions are equivalent in terms of measurement conditions such as the particle density, such that it is sufficient at this stage to not explicitly perform all studies for both positions. 
The main challenge of an electromagnetic calorimeter in this region of phase space is the requirement to discriminate decay photons from direct photons at very high energy, which will require extremely high granularity.

The design option currently under study is a SiW sandwich construction. It consists of 20 layers of a 3.5~mm W plate ($\approx 1 X_0$) interleaved with active layers with Si sensors. The active layers use two different sensor technologies: low granularity layers (LGL), which consist of sensors with 1 cm$^2$ pads summed longitudinally in segments and equipped with analog readout, and high granularity layers (HGL) based on CMOS monolithic active pixel sensors (MAPS). The MAPS will have a pixel size of a few $10 \times 10 \, \mu\mathrm{m}^2$ with internal binary readout.\footnote{The current model for MC simulations uses a pixel size of  $100 \times 100 \, \mu\mathrm{m}^2$.} On-chip processing will convert the pixel count in a \textit{macro-pixel} of $1 \, \mathrm{mm}^2$ to a pseudo-analog value.

\begin{figure}[htb]
 \begin{center}
    \includegraphics[width=0.45\textwidth]{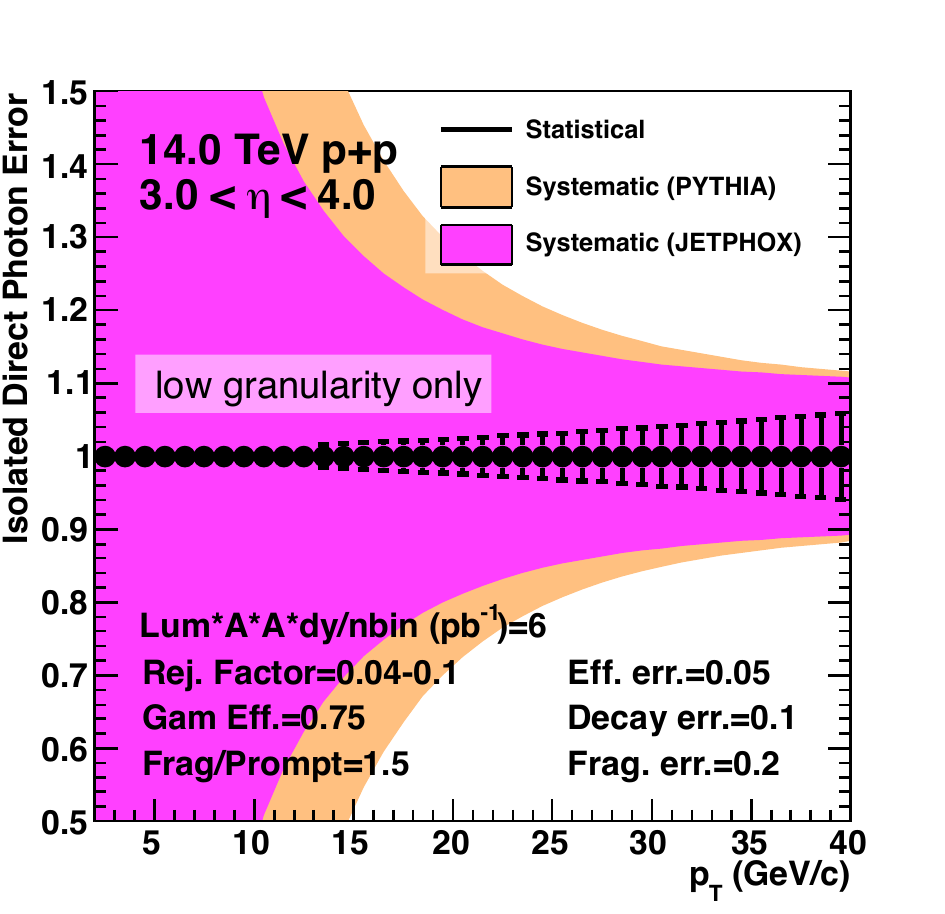}
    \includegraphics[width=0.45\textwidth]{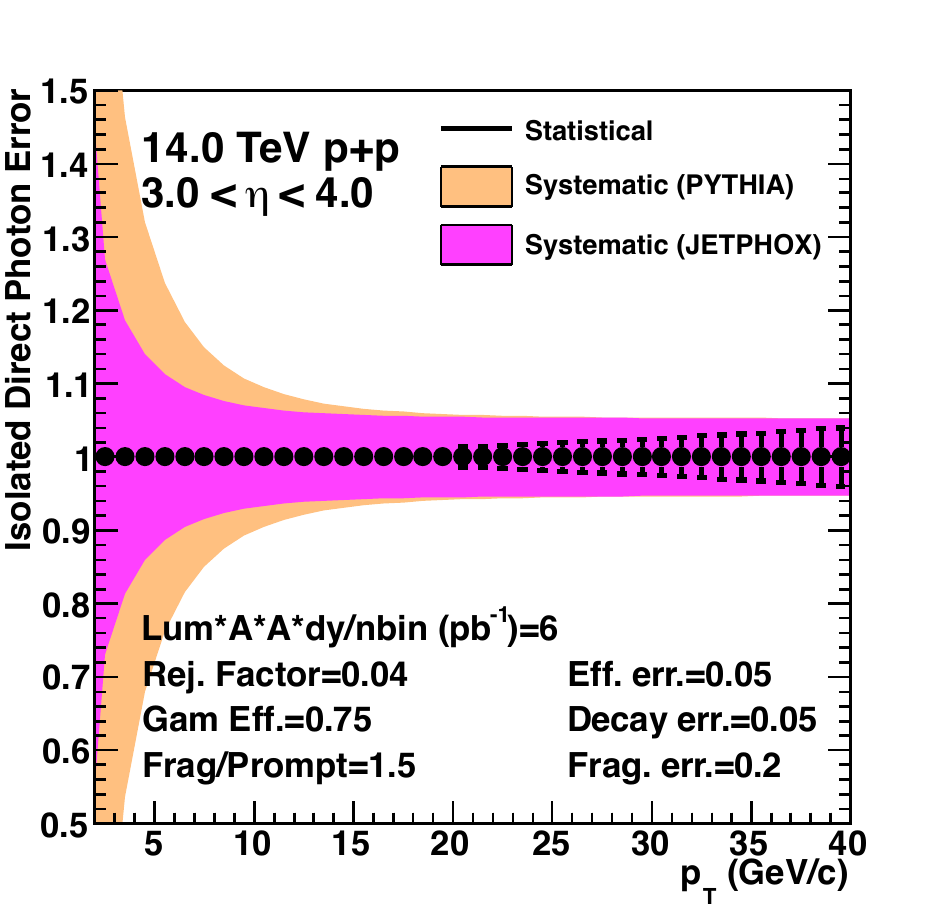}
   \caption{Estimated relative uncertainties on
  the cross section measurement for direct photon production in p+p collisions at $\sqrt{s}=14$
  TeV, based on direct photon spectra from \JETPHOX (dark band)
  and \PYTHIA (light band), and background spectra from \PYTHIA events. Statistical uncertainties are shown
  as error bars and the systematic uncertainty is shown as a band. This simulation assumes a location of the detector at $z = 3.6 \, \mathrm{m}$. Results are shown using only low granularity layers ({\it left}) and for the full detector including high granularity layers ({\it right}).
         }
   \label{fig-errors}
 \end{center}
\end{figure}

The HGL are crucial for  $\gamma/\pi^0$ discrimination. The LGL have an effective tower width of the order of the Moli\`ere radius. Their two-shower separation power is similar to existing standard electromagnetic calorimeters.\footnote{Those conditions are in fact very similar to the ones of the electromagnetic calorimeter of the LHCb experiment.} However, the shape of electromagnetic showers allows us to make use of much finer granularity for shower separation and additional shower shape analysis for very high energy $\pi^0$, when the two photons can no longer be resolved. The impact of the granularity is shown in Fig.~\ref{fig-errors}, which shows uncertainty estimates for a direct photon measurement in pp collisions at 14~TeV. The panel on the left hand side shows the expected performance using only the LGL, while the right panel shows the performance of the full detector. A low-granularity detector would only determine the photon yield with a much larger systematic error, mainly due to the merging of $\pi^0$-decay photons. Only the high-granularity option has a good  sensitivity for such a photon measurement. While FoCal would offer coverage towards higher rapidities than other LHC experiments, it is in particular the superior granularity at these large rapidities that would give FoCal a unique advantage.

\section{FoCal Prototype}

Within the R\&D program related to FoCal particular emphasis is put on the HGL technology. A full pixel calorimeter prototype has been designed and  constructed to validate the concept of a calorimeter using high-granularity CMOS pixel sensors with digital readout. The prototype has a quadratic cross section of $4 \times 4 \, \mathrm{cm}^2$ and a total depth of $\approx 10 \, \mathrm{cm}$. It consists of 24 layers of tungsten (3.3~mm) and sensor+infrastructure ($\approx 1 \, \mathrm{mm}$). This highly compact design leads to effective detector parameters of $X_0 \approx 4 \, \mathrm{mm}$ and $R_M \approx 11 \, \mathrm{mm}$.

As sensor we use only the MIMOSA23 chip \cite{Turchetta:tc} in this prototype. Chips of the MIMOSA series are one of the options for the final detector design and are also being investigated as a main solution for the ALICE ITS upgrade \cite{ITS:2012}. The active area of the MIMOSA23 consists of $640 \times 640$ pixels of a surface of $30 \times 30 \, \mathrm{\mu m}^2$ each. Every active layer uses $2 \times 2$ chips, providing $\approx 40$ million pixels in total. On one side of the chip a wide strip contains the row drivers, on the bottom a similar strip houses discriminators for every column, further readout logic and connections for input and output. Per layer partial overlap of the chips compensates for these non-active areas. Every chip is readout in a cycle of 640 $\mu \mathrm{s}$ by a so-called rolling shutter. 
The chips are wire-bonded to individual PCBs, which are then readout by a system of Spartan6 and Virtex6 FPGAs to a DAQ computer \cite{2013JInst...8P3015F}. No zero suppression is performed in the prototype, such that all pixels are read out continuously leading to an enormous data volume, which limits the rate capabilities. This will be different for the application of this technology in the FoCal detector, which will use an appropriate zero suppression and possibly further data reduction. All chips are connected to a cooling system via the tungsten plates. The prototype setup is complemented with a set of small scintillators for particle triggers.

\begin{figure}[bth] 
\begin{center}
\includegraphics[width=.8\textwidth]{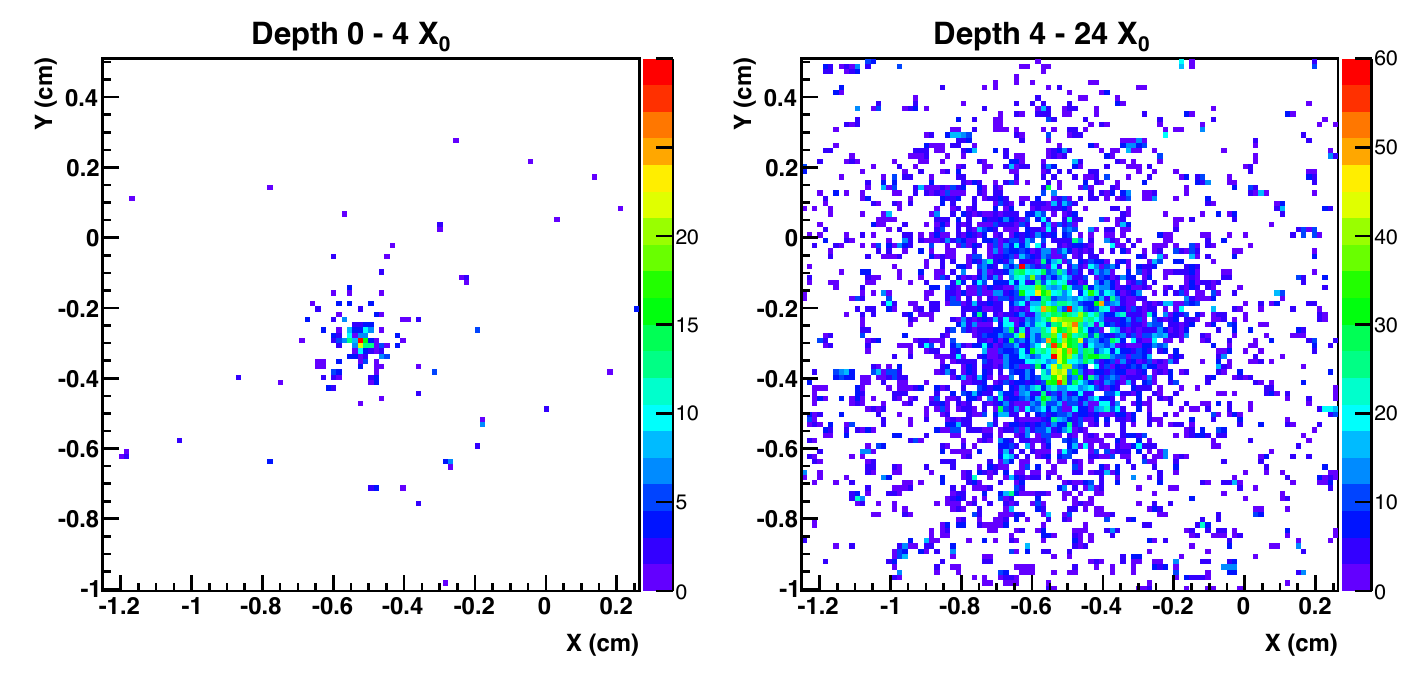}
\caption{Transverse density distribution of pixel hits in two different longitudinal sections of the FoCal prototype as measured for a single electron of 200 GeV.} 
\label{fig-shower} 
\end{center}
\end{figure}

\section{Results}

Test beam measurements have been performed at DESY with electron beams (2 and 5 GeV), and with mixed beams at the CERN PS (2 - 8 GeV) and SPS (30 - 250 GeV). The prototype has in addition been exposed to cosmic muons for an extended period of time. Fig.~\ref{fig-shower} shows a single event display in the transverse plane for an electron shower of 200~GeV. The left panel shows the hit density integrated over the first four layers, while the right panel shows the integrated hit density in the remaining layers. It can clearly be seen how the shower develops from a narrow to a broad distribution, as expected. Also, the narrow inner core of the shower ($\approx 1$~mm) is always apparent. The event display also testifies the low noise level of the detector. 

\begin{figure}[bth] 
\begin{center}
\includegraphics[width=.5\textwidth]{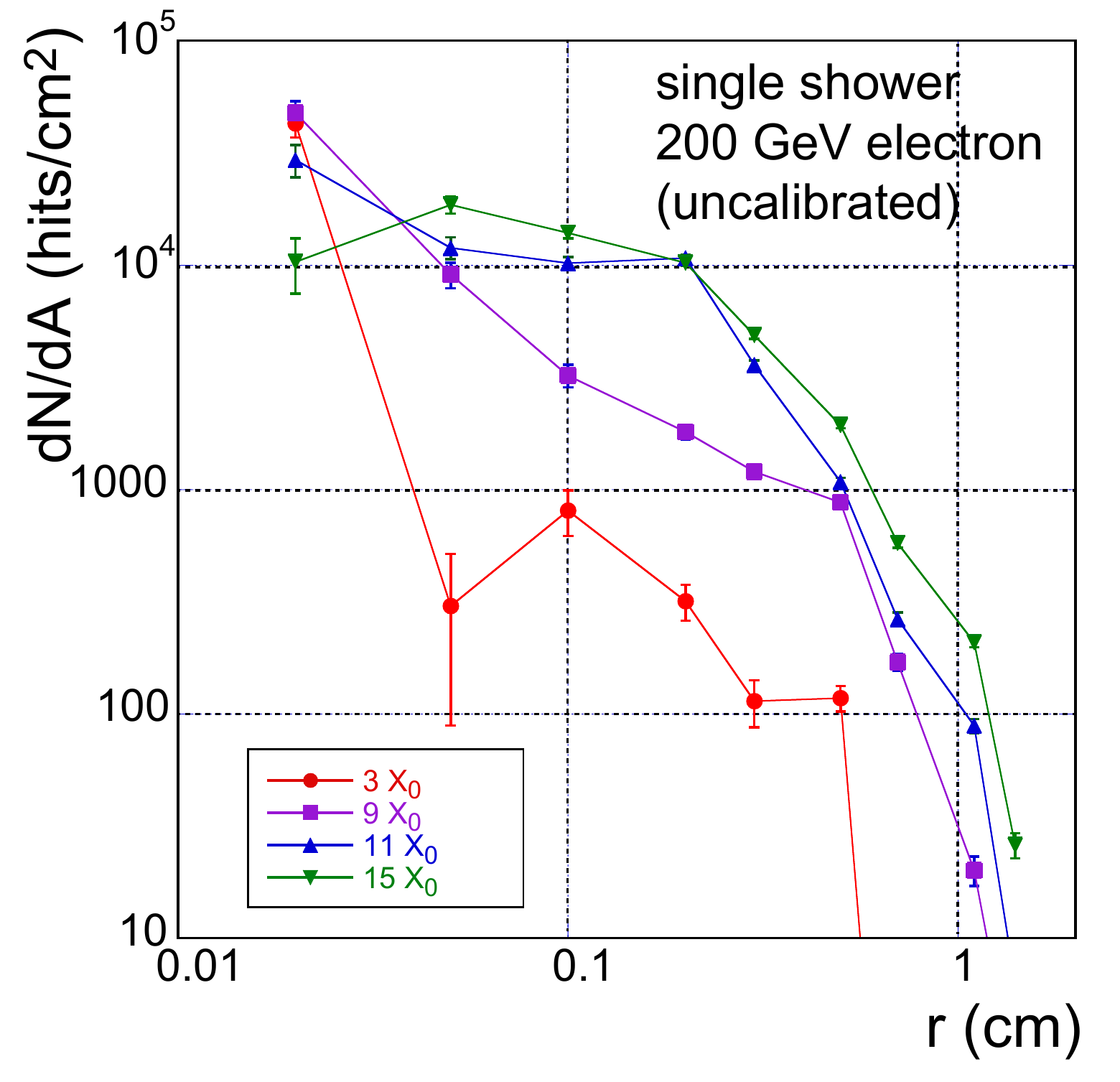}
\caption{Uncorrected hit distribution in single layers as a function of the transverse distance from the presumed shower centre for a single electron of 200 GeV.} 
\label{fig-dist} 
\end{center}
\end{figure}

This high granularity device can in fact give information of unprecedented detail on the shape of single showers on the sub-millimeter scale. Although the inter-calibration of chips and the alignment (on the scale of a pixel size) have not been finalised, this can already be demonstrated semi-quantitatively in Fig.~\ref{fig-dist}, which shows the transverse hit distributions in single layers as a function of the radial distance from the estimated shower centre, again for a single shower. The development from a very sharp peak in the early layer to a rather broad distribution deep in the detector is clearly seen. This gives confidence that a two-shower separation on the scale of a few mm should be achievable. The lateral profiles are also in good agreement with the estimated Moli\`ere radius of $R_M = 11 \, \mathrm {mm}$. Final results for these profiles will provide a wealth of information useful for particle identification and the development of shower separation algorithms.

\begin{figure}[t!] 
\begin{center}
\includegraphics[width=.8\textwidth]{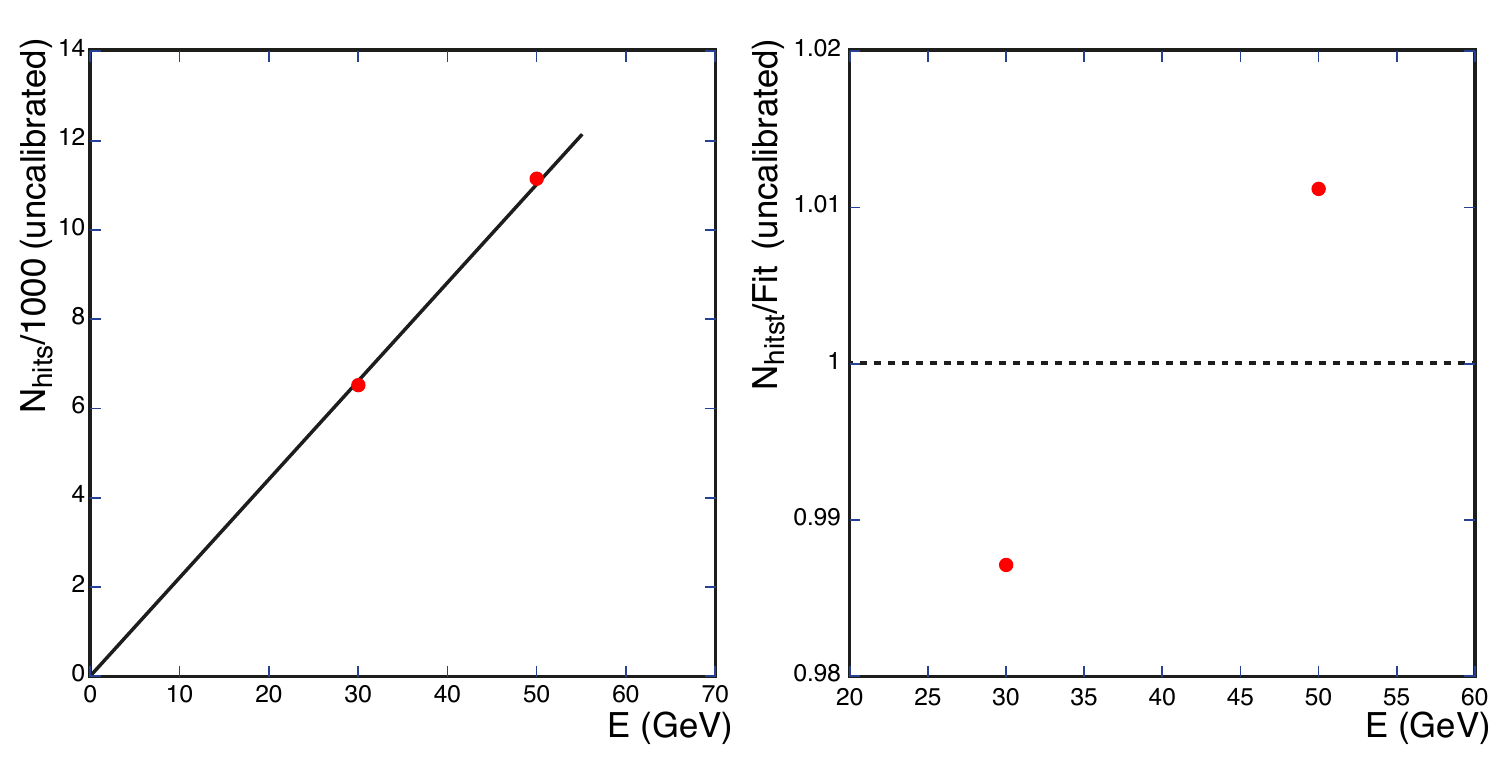}
\caption{Uncorrected detector response to electrons as a function of energy ({\it left}).  A fit of a linear function is also shown.  Ratio of the detector response to the value of the linear fit ({\it right}).} 
\label{fig-linear} 
\end{center}
\end{figure}

For the ALICE FoCal proposed the energy measurement would mainly rely on the LGL. Anyhow, for a forward measurement, where photon energies are usually very high, a satisfactory energy resolution is relatively easy to achieve. However, the relative sharing of energy between two photons merging in the LGL will be estimated from the energy information of the HGL, and the latter is therefore important for the $\pi^0$ reconstruction efficiency at high energy. 
In addition, the prototype shall also serve as a proof-of-principle for digital calorimetry, i.e. to measure the deposited energy of an electromagnetic shower via the number of hits from charged particles in the sensitive layers. As such, also the linearity and resolution of the prototype is naturally of interest. For this purpose the calibration of the detector is crucial, but at the same time it is a non-trivial task for such a device: It is impossible to perform a calibration on the pixel level, while a single calibration factor for every chip may not be sufficient in all cases. Calibration will likely have to be performed on an intermediate scale, and appropriate algorithms are still being developed. In addition, a number of chips are either very noisy or give no signal, and this will strongly effect the resolution obtained. Taking into account these current limitations a reasonable linearity is obtained for data sets taken under similar conditions. Fig.~\ref{fig-linear} shows the detector response for two energies without any correction applied. In the left panel the data points are shown together with a linear fit, the right panel shows the ratio of the data to this fit. Statistical errors are negligible for these results, but it is obvious that a systematic correction is necessary. Even without any correction the detector response already agrees with the proportionality within $\approx 1\%$.

\begin{figure}[t!] 
\begin{center}
\includegraphics[width=.9\textwidth]{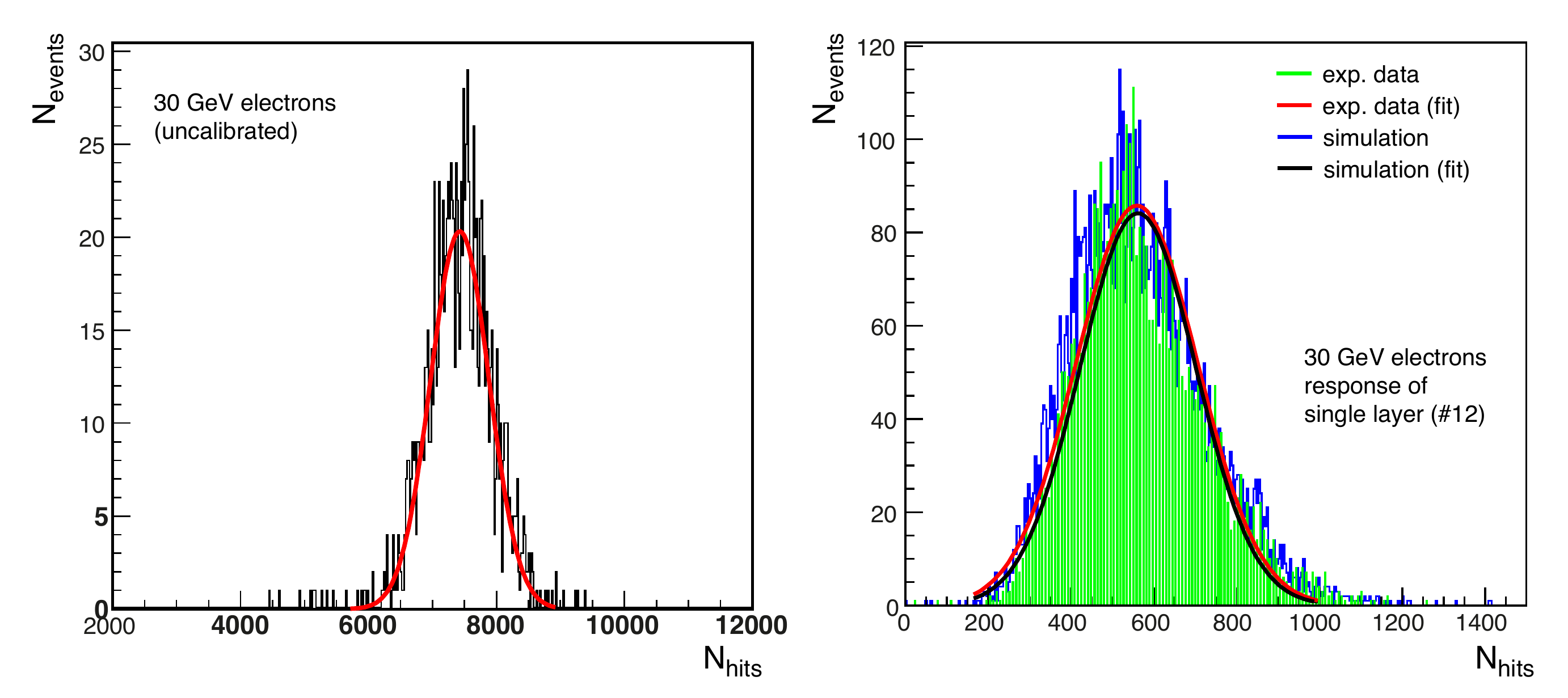}
\caption{ Spectrum of uncorrected hit response of the full prototype to 30 GeV electrons ({\it left}). Response of layer 12 for 30 GeV ({\it right}). Here the experimental data are compared to \GEANT simulations, where the amplitude is scaled to match the experimental data.} 
\label{fig-resolution} 
\end{center}
\end{figure}

The detector response spectrum to electrons obtained without calibration is shown in Fig.~\ref{fig-resolution}. The response of the full detector as shown in the left panel yields a resolution of $\sigma = 6.2 \%$ at 30 GeV (as obtained with a Gaussian fit), which is about a factor of 2 larger than the expectation from \GEANT simulations. Dead chips have been taken into account in the simulations, so the discrepancy is most likely due to missing calibration corrections. To reduce the effect of possible missing calibration we have also studied the single-layer response, which is to a large extent dominated by the performance of single chips. Fig.\ref{fig-resolution} shows on the right the hit spectrum of layer 12. It is compared to \GEANT calculations, which are appropriately scaled  -- both are shown with corresponding Gaussian fits. For this single-layer response the measurement shows good agreement with the simulation. For the experimental data we obtain $\sigma_{exp} = 26.3 \%$, while the simulation yields  $\sigma_{sim} = 25.0 \%$.

\section{Summary}
A prototype of a very-high granularity SiW-calorimeter for the ALICE FoCal upgrade proposal has been built and successfully used in first beam tests. The very compact design of this detector has allowed us to obtain a Moli\`ere radius as small as $R_M = 11 \, \mathrm {mm}$. A large fraction of the 39 million MIMOSA23 pixels are read out, and data have been taken for a wide range of energies. While calibration of the detector is still ongoing, encouraging results are obtained for linearity and resolution. The detector has unprecedented capabilities for shower shape analysis and two-shower separation. As the proposed FoCal detector in its current design makes use of only two single HGL, the single layer response shown here indicates already that the technology should be adequate.

\bibliography{mybib}{}
\bibliographystyle{unsrt}

\end{document}